\documentclass[prd,preprint]{revtex4}
\usepackage[english]{babel}
\usepackage{amscd}
\usepackage{epsfig}
\usepackage{tabularx}
\usepackage{graphicx}
\usepackage{latexsym}
\usepackage{amsmath}
\usepackage{amsfonts}
\usepackage{amssymb}

\usepackage{orcidlink}
\usepackage{indentfirst}
\usepackage{hyperref}
\usepackage{times}
\usepackage[T1]{fontenc}
\usepackage{latexsym}
\usepackage{graphics}
\usepackage{verbatim}
\usepackage[absolute]{textpos}
\usepackage{wrapfig}
\usepackage{amsthm}
\usepackage{setspace}
\usepackage{color}
\definecolor{Red}{rgb}{0.9,0.1,0.1}
\definecolor{blue}{rgb}{0.25,0.25,0.6}
\usepackage{booktabs, multirow, tabularx} \usepackage{pgfplots} \pgfplotsset{compat=1.18} \usepackage{xcolor} \usepackage{caption, subcaption} \usepackage{hyperref} \hypersetup{ colorlinks=true, linkcolor=blue!70!black, citecolor=blue!70!black, urlcolor=blue!70!black } \definecolor{blueheader}{RGB}{31, 78, 121} \definecolor{tempfora}{RGB}{41, 128, 185} \definecolor{tempdentro}{RGB}{192, 57, 43} \definecolor{deltacolor}{RGB}{39, 174, 96}

\newcommand{\be}{\begin{equation}}
	\newcommand{\ee}{ \end{equation}}
\newcommand{\ben}{\begin{eqnarray}}
	\newcommand{\een}{\end{eqnarray}}

\begin{document}
	
	\title{From Local Context to Climate Science: Low-Cost Experimentation in the Initial Training of Physics Teachers}
	
	\author{André A. A. Marinho$^{1,*}$, Edson R. L. Lopes$^{1}$, Jhon KF V. Silva$^{1}$, Mariana S. Costa$^{1}$, Maria A. Morais$^{2}$}
	
	\affiliation{$^1$Centro de Ciências Exatas, Naturais e Tecnológicas - CCENT, Universidade Estadual da Região Tocantina do Maranhão (UEMASUL), 65901-480, Imperatriz, MA, Brazil\\
		$^2$Instituto Federal do Maranhão - IFMA, Campus Açailândia, 65930-000, Açailândia, MA, Brazil}
	
	
	
	\email{andre.afonso@uemasul.edu.br}

	\begin{abstract}
		Climate change poses a major scientific, environmental, and educational challenge, requiring approaches that connect physical concepts, empirical evidence, and socially relevant environmental issues. This study presents and analyzes a low-cost experimental investigation developed within the "Physics and the Environment" course of a Physics teacher training program at the State University of the Tocantina Region of Maranhão (UEMASUL), Brazil. The experiment involved constructing a simplified thermal simulator using accessible materials and comparing a confined system with an open control system, both exposed simultaneously to solar radiation. The results showed that the average temperature inside the simulator was $45.91^{\circ}\mathrm{C}$, compared to $38.58^{\circ}\mathrm{C}$ in the external control system, corresponding to a mean temperature difference of $7.33^{\circ}\mathrm{C}$ (among other analyses). These results demonstrate that the experimental setup was capable of producing and measuring a significant thermal difference between the two systems. However, the observed temperature rise should not be interpreted as a direct measure of the atmospheric greenhouse effect, given that the experimental apparatus involves both radiative and non-radiative heat transfer mechanisms. The study emphasizes the educational potential of low-cost experimentation as a strategy to integrate electromagnetic radiation, energy balance, temperature, heat transfer, and climate science into the initial training of physics teachers. It also highlights the importance of explicitly discussing the limitations of experimental models and distinguishing local experimental evidence from conclusions regarding global climate processes.

	\end{abstract}
	\maketitle
	\section{INTRODUCTION}
	
	Observed changes in the climate system constitute one of the most significant scientific, environmental, and educational challenges of the twenty-first century. Understanding these phenomena requires moving beyond interpretations based exclusively on temperature variations or isolated weather events and considering Earth as a dynamic system in which the atmosphere, hydrosphere, cryosphere, biosphere, and land surface interact through complex flows of matter and energy. In this context, fundamental concepts of physics - such as radiation, temperature, energy transfer, thermal equilibrium, and conservation laws - play a central role in understanding the processes that determine the state and evolution of the climate system. Earth's energy balance results from the relationship between incoming solar radiation, the reflected portion, and the thermal radiation emitted by the Earth-atmosphere system; consequently, changes in these fluxes can trigger thermal and climatic responses across various temporal and spatial scales.\cite{forster, ipcc}.
	
	The importance of this knowledge has become even more evident since the Industrial Revolution, when the intensification of fossil-fuel use and industrial activities produced significant changes in atmospheric composition \cite{crutzen}. The anthropogenic increase in greenhouse-gas concentrations has altered the planet's radiative balance and is the main factor responsible for the global warming observed in recent decades \cite{ipcc}. However, understanding this process requires a fundamental conceptual distinction: the greenhouse effect is a natural physical phenomenon essential for maintaining thermal conditions compatible with life on Earth, whereas global warming is largely associated with the anthropogenic intensification of this effect \cite{khmelinskii}. This distinction is particularly relevant in science education, since simplified interpretations may lead to misconceptions about the relationships among radiation, the atmosphere, and temperature \cite{junges, ipcc}.
	
	From a physical standpoint, the greenhouse effect is related to the spectral properties of matter and the interaction of electromagnetic radiation with atmospheric constituents. Radiation from the Sun reaches the Earth system predominantly at wavelengths associated with shortwave radiation, including the visible and near-infrared regions \cite{pierrehumbert}. Part of this energy is reflected, while another fraction is absorbed by the atmosphere and the surface. The heated Earth's surface, in turn, emits radiation predominantly in the thermal infrared. Molecules present in the atmosphere, such as carbon dioxide, water vapor, and methane, interact with specific ranges of this radiation through absorption and emission processes, contributing to radiative exchanges among the Earth's surface, the atmosphere, and space \cite{forster, ipcc}.
	
	The energy associated with electromagnetic radiation can be expressed by the relation
	
	\begin{equation}
		E=h\nu=\frac{hc}{\lambda},
	\end{equation}
	where $E$ represents energy, $h$ is Planck's constant, $\nu$ is frequency, $c$ is the speed of light, and $\lambda$ is wavelength. This relationship makes it possible to understand that different regions of the electromagnetic spectrum have different associated energies and that the interaction between radiation and matter depends on the properties of the radiation and the material constituents involved.
	
	The energy balance of the Earth system can be represented, in a simplified manner, by
	\begin{equation}
		\frac{dE}{dt}=R_{\mathrm{in}}-R_{\mathrm{out}},
	\end{equation}
	where $E$ represents the energy stored in the system, $R_{\mathrm{in}}$ is the incoming energy flux, and $R_{\mathrm{out}}$ is the emitted energy flux. Under an idealized condition of energy equilibrium,
	
	\begin{equation}
		R_{\mathrm{in}}=R_{\mathrm{out}}.
	\end{equation}
	
	When a persistent difference occurs between these fluxes, the system begins to accumulate or lose energy, producing responses in the thermal state of the climate system \cite{forster}. Although this representation is simplified and does not encompass the full complexity of climate, it provides an important physical basis for discussing the relationship among radiation, energy, and temperature.
	
	Understanding this phenomenon therefore involves more than observing that a given environment becomes warmer when exposed to solar radiation. It is necessary to understand the mechanisms of radiation absorption, emission, and reflection, as well as the energy-transfer processes that act simultaneously \cite{landulfo}. In the climate system, energy is transferred through radiative and non-radiative processes, including conduction, convection, and latent heat transfer. The joint analysis of these mechanisms makes it possible to establish a direct connection between topics traditionally addressed in Physics and environmental phenomena observable in everyday life \cite{forster}.
	
This relationship is particularly relevant in the context of the Physics and Environment course, especially when incorporated into Physics teacher education. Contemporary teacher education requires more than isolated mastery of disciplinary concepts; it requires future teachers to relate scientific knowledge, environmental problems, and concrete learning situations \cite{magalhaes, reis}. Recent literature highlights the need to prepare teachers to address climate change in a scientifically grounded, interdisciplinary manner and in connection with local contexts, while also fostering students' ability to critically interpret environmental problems \cite{beach, singh, ibourk}.

In this sense, scientific experimentation constitutes a particularly relevant possibility. Educational experiments make it possible to bring abstract concepts closer to observable phenomena, encouraging the formulation of hypotheses, the collection and interpretation of data, and the construction of explanations grounded in evidence. Specifically in relation to the greenhouse effect, low-cost experiments have been proposed as strategies for discussing the interaction of infrared radiation with greenhouse gases and for bringing physical concepts closer to experimental situations that are accessible to students \cite{junges}.

However, the use of accessible materials may serve a purpose that goes beyond simply reducing costs. In educational contexts characterized by limitations in laboratory infrastructure, experimental devices constructed from easily obtainable materials can expand opportunities for scientific experimentation. When these devices are developed, constructed, and analyzed by preservice teachers themselves, experimentation simultaneously becomes a strategy for learning Physics and a teacher-training experience for their future professional practice.

This perspective is particularly important for the scientific education of children and adolescents. Contemporary environmental issues, such as climate change, global warming, and changes in atmospheric conditions, require educational practices that enable students to interpret scientific phenomena based on evidence and establish relationships among science, the environment, and society \cite{landulfo}. Teacher education is a fundamental element in this process, since difficulties related to scientific knowledge, confidence in teaching climate change, and the articulation among different fields of knowledge may limit the treatment of this topic in basic education \cite{beach, ibourk}.

Climate-related science education may also contribute to an educational process that is not limited to the transmission of information but instead promotes an understanding of physical processes, evidence analysis, and responsible participation in problems that affect present and future living conditions. In this sense, scientific education also assumes a social dimension, since children and adolescents will be directly involved in decisions and transformations related to the environment over the coming decades \cite{singh}.

In this context, using a global environmental phenomenon as the starting point for a locally situated investigation presents significant pedagogical potential. The present experiment was developed at the Universidade Estadual da Região Tocantina do Maranhão (UEMASUL), located in the city of Imperatriz, Maranhão, Brazil, at $5^{\circ}33'40.3''$ S and $47^{\circ}28'49.0''$ W (see Fig.~\ref{1}). The investigation was based on the construction of a low-cost experimental device designed to investigate heating processes associated with exposure to solar radiation. Data collection was carried out over eleven days, covering five days before and five days after the June 2026 solstice. The solstice was used as a temporal and astronomical reference to delimit the investigation and not as an isolated causal factor to explain the observed temperature variations.

It is important to emphasize that an eleven-day time series cannot characterize the climate of a region or establish a global warming trend. Climate involves statistical patterns and behaviors analyzed over much longer time scales, whereas short-duration observations are more directly related to weather conditions and short-term variability. This distinction is fundamental to avoid interpretations that exceed the limits of the experimental data. At the same time, short time series can constitute important educational resources for discussing variability, fluctuations, and the dependence of physical processes on environmental conditions \cite{varley}. The observation of different temperature values over a specific period makes it possible to discuss the fact that environmental systems are not static and that their variables result from the simultaneous action of different mechanisms.

The variations observed in a temperature time series can therefore serve as a starting point for a broader discussion of the complexity of the climate system. Although a short-duration educational experiment cannot characterize the climate of a region or infer global trends, its data can demonstrate that environmental variables fluctuate and that their interpretation requires consideration of different processes and scales. Climate dynamics exhibit nonlinear behavior, interactions among processes at different scales and the simultaneous influence of deterministic and stochastic mechanisms \cite{ghil, lucarini}. In this context, concepts related to chaos theory, scales, and fractal structures can be incorporated as conceptual tools for discussing the complexity of environmental variables, without assigning these concepts a direct causal role in explaining the greenhouse effect \cite{lopes, azevedo, mandelbrot}. The atmosphere constitutes a system in which phenomena occur simultaneously at different spatial and temporal scales, exhibiting characteristics associated with variability, turbulence, nonlinearity, and complex structures.

In the present study, chaos theory is not used to directly explain the greenhouse effect or to claim that the experimental time series exhibits chaotic behavior. Its use is conceptual and pedagogical, allowing discussion of how environmental phenomena may exhibit variability, nonlinear relationships, and interactions among different scales. Likewise, fluctuations in a short temperature series do not constitute evidence of deterministic chaos. Demonstrating such behavior would require longer time series and specific mathematical and statistical procedures.

Against this background, this study presents and analyzes an experimental investigation developed within the Physics and Environment course of an undergraduate Physics teacher education program, in which students constructed a low-cost experimental device intended for the didactic investigation of processes related to the greenhouse effect and radiative heating. The proposal sought to articulate concepts of electromagnetic radiation, energy balance, temperature, and heat transfer with issues related to climate change and science education.

The investigation therefore adopts a perspective that considers experimentation not merely as a demonstrative procedure but as a strategy for scientific and teacher education. By involving future teachers in the construction, observation, recording, and interpretation of an environmental phenomenon, the study seeks to contribute to the development of competencies that may subsequently be mobilized in the science education of children and adolescents. The study also seeks to discuss the possibilities and limitations of short-duration, low-cost experiments for the didactic treatment of complex climate phenomena, emphasizing the need to distinguish local experimental evidence from interpretations concerning global climate processes.
	
	\section{THEORETICAL FRAMEWORK}
	\subsection{Greenhouse Effect, Solar Spectrum, and Interaction of Radiation with Matter}

	The Earth's climate system can be understood from a physical perspective through the energy exchanges occurring among the Sun, atmosphere, Earth's surface, and outer space. The energy balance at the top of the atmosphere represents the relationship between the energy entering the system and the energy leaving it. When this balance is altered by natural or anthropogenic forcings, the climate system responds through changes in its energy content and thermal state \cite{forster}.
	
	In a simplified representation, the average solar radiation absorbed by the Earth system can be expressed as
	
	\begin{equation}
		R_{\mathrm{solar}}=(1-\alpha)\frac{S_0}{4},
	\end{equation}
	where $S_0$ represents the average solar irradiance at the top of the atmosphere and $\alpha$ represents planetary albedo. The factor $(1/4)$ is related to the ratio between the Earth's cross-sectional area, which intercepts solar radiation, and its total surface area.
	
	The emitted thermal radiation can be approximately related to temperature through the Stefan–Boltzmann law,
	
	\begin{equation}
		R_{\mathrm{thermal}}=\sigma T^4,
	\end{equation}
	where $\sigma$ is the Stefan–Boltzmann constant and $T$ is the absolute temperature.
	
	Thus, in an extremely simplified radiative model, an equilibrium condition can be represented by
	
	\begin{equation}
		(1-\alpha)\frac{S_0}{4}=\sigma T^4.
	\end{equation}
	
	This expression does not represent a complete model of Earth's climate because it does not explicitly describe atmospheric absorption, clouds, convection, water vapor, energy storage by the oceans, or other climate processes and feedbacks. Its importance lies in presenting an elementary physical relationship between absorbed solar energy and emitted thermal radiation.
	
	The greenhouse effect modifies this simplified representation because the atmosphere interacts with infrared radiation emitted by the Earth's surface. Greenhouse gases absorb and emit radiation in specific regions of the spectrum, altering how energy is transferred among the surface, atmosphere, and outer space. The natural greenhouse effect is therefore a fundamental component of Earth's energy balance, whereas the anthropogenic increase in greenhouse gas concentrations modifies the radiative forcing of the climate system \cite{ipcc}.
	
	The distinction between the greenhouse effect and global warming is therefore essential. The former corresponds to a natural physical process associated with the radiative properties of the atmosphere; the latter corresponds to the persistent increase in the Earth's mean surface temperature observed over long time scales and predominantly associated with human influences on the climate system.
	
	For an educational experimental activity, this distinction has methodological importance. A device constructed from low-cost materials does not reproduce the Earth's atmosphere on a reduced scale and should not be interpreted as a complete reproduction of the planetary greenhouse effect. Its role is to enable the investigation of specific physical processes related to radiation, energy absorption, heating, and thermal exchanges \cite{junges}.
	
	Solar radiation comprises a broad range of the electromagnetic spectrum \cite{young}. Its spectral distribution is fundamental for understanding the energy received by Earth and the subsequent thermal response of the surface and atmosphere.
	
	The energy associated with a photon of electromagnetic radiation is given by
	
	\begin{equation}
		E=\frac{hc}{\lambda}.
	\end{equation}
	
	Thus, radiation with shorter wavelengths has greater energy per photon, whereas radiation with longer wavelengths has lower energy per photon. However, the interaction between radiation and matter depends not only on radiation energy but also on the structural and spectral properties of the material.
	
	The Earth's surface receives predominantly shortwave solar radiation and, after being heated, emits predominantly longwave thermal radiation. This distinction is fundamental for understanding the greenhouse effect because atmospheric constituents exhibit selective absorption and emission properties in the infrared region \cite{forster, ipcc}.
	
	This relationship makes it possible to integrate topics that are often treated separately in Physics education, such as electromagnetic waves, the electromagnetic spectrum, radiation, radiation–matter interactions, temperature, and thermodynamics. Experimentation can help make these relationships more concrete by allowing thermal responses to be observed and the associated physical processes to be discussed.
	
	\subsection{Thermal Gradients, Energy Transfer, and Energy Conservation}
	
	Temperature is a fundamental variable for characterizing the thermal state of a system. In environmental systems, however, its interpretation should not be limited to the absolute value measured at a particular instant. The spatial and temporal distribution of temperature is also relevant because temperature differences are associated with energy transfer and, in fluid systems, may contribute to mass motion.
	
	The temperature gradient measures the rate and direction in which temperature changes in space. It is fundamental for understanding the movement of air masses, wind formation, and atmospheric instability as altitude changes. It can be represented by $\nabla T$, a vector quantity indicating the direction of greatest spatial temperature variation.
	
	In thermal conduction processes, the heat flux can be described, in its simplest form, by Fourier's law:
	
	\begin{equation}
		\vec{q}=-\kappa\nabla T,
	\end{equation}
	where $\vec{q}$ represents the heat flux, $\kappa$ is the thermal conductivity of the material, and $\nabla T$ is the temperature gradient.
	
	The negative sign indicates that heat flows spontaneously toward decreasing temperatures. Although the experiment developed in this study does not aim to isolate conduction from the other energy-transfer mechanisms, this relationship is useful for establishing the physical meaning of the observed temperature differences.
	
	In the climate system, however, energy transfer also occurs through convection, radiation, and latent heat transport. Therefore, the observed temperature variations should be understood as the result of the simultaneous action of different physical processes.
	
	The physical description of the climate system is also based on conservation laws. Conservation of energy establishes that energy is neither created nor destroyed but can be transferred and transformed between different forms. In a simplified representation,
	
	\begin{equation}
		\frac{dE}{dt}=P_{\mathrm{in}}-P_{\mathrm{out}},
	\end{equation}
	where $P_{\mathrm{in}}$ and $P_{\mathrm{out}}$ represent the rates of incoming and outgoing energy, respectively.
	
	In the Earth system, solar energy can be reflected, absorbed, transformed into thermal energy, transported by atmospheric and oceanic motions, or associated with phase changes of water. These processes occur simultaneously and contribute to the evolution of the energy state of the climate system.
	
	Mass conservation is also fundamental to the description of the atmosphere. Air motion involves the transport of mass, water vapor, aerosols, and other constituents. In general terms, the continuity equation can be expressed as
	
	\begin{equation}
		\frac{\partial \rho}{\partial t}+\nabla\cdot(\rho\vec{v})=0,
	\end{equation}
	where $\rho$ represents fluid density and $\vec{v}$ its velocity field.
	
	Likewise, conservation of momentum constitutes the basis for describing atmospheric fluid motion. In simplified form,
	
	\begin{equation}
		\rho\frac{D\vec{v}}{Dt}=\vec{F},
	\end{equation}
	where $D/Dt$ represents the material derivative and $\vec{F}$ represents the forces acting on the fluid.
	
	These equations are not directly solved in the experiment. Their inclusion in the theoretical framework aims to demonstrate that the climate system cannot be reduced to temperature alone. The evolution of environmental variables is simultaneously related to energy fluxes, mass transport, fluid motion, radiation, and phase changes.
	
	\subsection{Chaos Theory and Environmental Phenomena}
	
	A central issue in interpreting environmental data is the distinction between weather, climate variability, and climate change. Weather is associated with atmospheric conditions observed over relatively short periods, whereas climate involves statistical patterns analyzed over longer time scales. Climate change refers to persistent changes in the characteristics of the climate system.
	
	This distinction has direct implications for interpreting the results of this study. An eleven-day experimental series can reveal temperature variations during the investigated period, but it cannot establish a regional or global climate trend. Its scientific and educational value lies in the possibility of analyzing the observed variations, formulating hypotheses about the physical mechanisms involved, and discussing the limits of inferences based on the data.
	
	Climate science depends on the analysis of observations, physical models, and time series developed across different spatial and temporal scales. Internal variability of the climate system interacts with different forcings, making its interpretation dependent on considering multiple processes and scales \cite{forster, ghil}.
	
	The climate system exhibits characteristics associated with complex dynamical systems. Its different components interact continuously, and the resulting responses cannot always be described by simple linear relationships. Distinct processes may occur simultaneously at different spatial and temporal scales, while changes in external conditions may modify the evolution of the system. Dynamical systems theory and studies related to chaos have contributed to the development of approaches aimed at understanding atmospheric and climate variability. Climate dynamics involve nonlinear interactions, internal variability, transitions between different regimes, and predictability limits, making it necessary to consider the temporal evolution of the system and interactions among its components \cite{ghil, lucarini}.
	
	In this study, chaos theory is introduced as a conceptual and pedagogical reference. It is not intended to directly explain the greenhouse effect or to claim that the eleven-day experimental series exhibits chaotic behavior. Such a claim would require longer time series and specific mathematical and statistical analyses.
	
	The contribution of chaos theory to the proposed discussion therefore lies in the possibility of addressing concepts such as nonlinearity, variability, sensitivity, and interactions among different scales, allowing students to recognize that environmental phenomena may exhibit behaviors that cannot be adequately described by simple cause-and-effect relationships.
	
	The concept of fractals is related to structures that exhibit self-similarity or scale-related properties at different levels of observation \cite{mandelbrot}. Although not every complex pattern observed in nature is necessarily fractal, concepts related to scaling and multiscale structures have been employed in the study of atmospheric and geophysical phenomena.
	
	The atmosphere exhibits structures distributed from very small scales to large circulation patterns. This multiscale organization contributes to the complexity of atmospheric dynamics and makes the relationship between local observations and large-scale behavior an issue requiring caution and theoretical grounding \cite{ghil}.
	
	In the present study, the concepts of fractals and scales are used as references for discussing the multiscale nature of environmental phenomena. The study does not claim that the experimental data exhibit a fractal structure. Identifying fractal properties would require specific mathematical and statistical procedures that are beyond the objectives of this study.
	
	\subsection{Physics and Environment in Teacher Education}
	
	The incorporation of environmental issues into initial Physics teacher education expands the possibilities for articulating disciplinary knowledge with scientific and social problems \cite{landulfo}. The Physics and Environment course can constitute a space for integrating physical concepts, environmental phenomena, and pedagogical practices, contributing to more contextualized teacher education. Preparing teachers to address climate change involves challenges related to scientific knowledge, confidence in teaching the topic, interdisciplinarity, and the ability to establish relationships between global processes and local contexts \cite{beach}. Recent studies also indicate the importance of developing knowledge about climate change and teacher self-efficacy for teaching this topic \cite{ibourk}.
	
	In this context, the construction of experiments by preservice teachers themselves may promote teacher education that integrates conceptual knowledge, experimental investigation, and pedagogical planning. The future teacher ceases to be merely a user of a previously designed activity and instead experiences the stages of device construction, definition of experimental conditions, measurement, data analysis, and discussion of limitations.
	
	Experimentation plays an important role in science education because it makes it possible to bring theoretical concepts closer to observable phenomena. When associated with the formulation of questions, development of hypotheses, measurements, analysis of results, and discussion of limitations, it can contribute to the development of scientific reasoning.
	
	In the teaching of the greenhouse effect, low-cost experiments present a particularly relevant possibility. In \cite{junges}, the authors developed a low-cost experimental proposal designed to demonstrate the absorption of infrared radiation by carbon dioxide, highlighting the potential of accessible experimental activities for discussing concepts related to the greenhouse effect.
	
	The use of low-cost materials also has social and pedagogical relevance. In schools where fully equipped laboratories are unavailable, accessible materials can expand opportunities for experimental activities. However, low cost should not imply reduced scientific rigor. On the contrary, simple experimental systems require careful attention to boundary conditions, variables involved, uncertainties, and the limitations of the model being used.
	
	This perspective is important because science education does not consist merely of obtaining numerical results. It also involves model construction, evidence evaluation, comparison of explanations, hypothesis formulation, and recognition of the limitations of experimental procedures.
	
	Investigating environmental phenomena from local contexts can strengthen the relationship between scientific knowledge and everyday experiences. Temperature, solar radiation, energy transfer, and atmospheric conditions are phenomena that can be directly observed, making it possible to relate scientific concepts to concrete situations.
	
	This approach is particularly relevant to climate change education. Understanding a global phenomenon does not require ignoring local contexts; on the contrary, observations conducted in environments close to students may serve as starting points for discussing broader scientific processes. Teacher education can contribute to this connection by preparing future teachers to address environmental issues in a contextualized and scientifically rigorous manner \cite{beach, singh}.
	
	In the present study, conducting the experiment at $5^{\circ}33'40.3''$ S and $47^{\circ}28'49.0''$ W provides a concrete environmental context for the investigation. The local focus is not used to generalize the results to other regions. Its purpose is to bring the scientific investigation process closer to the reality in which students live and in which future teachers may develop their professional activities.
	
	\subsection{Articulation among Physics, Environment, and Scientific Education}
	
	The articulation among Physics, the environment, and teacher education constitutes the central axis of this proposal. The investigated phenomenon makes it possible to mobilize knowledge related to electromagnetic radiation, energy transfer, temperature, thermodynamics, and dynamical systems while simultaneously discussing climate change, environmental responsibility, and science education.
	
	The proposal also makes it possible to understand experimentation as an activity that goes beyond demonstration. By constructing and investigating an experimental device, preservice teachers develop an experience that combines scientific knowledge, investigative skills, and pedagogical knowledge. This integration may contribute to greater autonomy in designing contextualized experimental activities for basic education.
	
	The main contribution of this study therefore does not lie in using a short-duration experiment to draw conclusions about global climate behavior. Rather, it lies in demonstrating how a simple experimental investigation can serve as a starting point for discussing complex physical and environmental phenomena. Based on data produced in a local context, it becomes possible to discuss energy balance, radiation, temperature, variability, complexity, and the limits of scientific interpretation.
	
	Thus, the experiment developed within the Physics and Environment course is understood as a strategy for scientific and teacher education. Its use makes it possible to bring Physics closer to contemporary environmental issues while simultaneously creating conditions for future teachers to experience an investigative practice that can later be adapted for science education involving children and adolescents.

	\section{Methodology}
	
	The present study is characterized by a qualitative and quantitative experimental approach, combining the construction and application of a low-cost thermal simulator with scientific investigation, science communication, and socio-environmental awareness activities. The proposal was developed through the articulation of Physics knowledge and environmental issues, considering the possibilities of using phenomena related to climate change as contextual elements for Physics teaching \cite{reis}.
	
	The methodology was organized into four main stages: (i) characterization and theoretical foundation of the extension project; (ii) implementation of awareness and science communication activities; (iii) bibliographic review, adaptation, and construction of the experimental apparatus; and (iv) execution of the experimental procedure and systematic collection of temperature data.
	
	\subsection{Characterization of the Extension Project}
	
	The intervention was developed within an extension project at the Universidade Estadual da Região Tocantina do Maranhão (UEMASUL), linked to the undergraduate Physics teacher education program. The activities were structured around the articulation of scientific knowledge, environmental issues, and situations related to everyday life.
	
	\begin{figure}[htb!]
		\includegraphics[{angle=90,height=5.0cm,angle=270,width=5.0cm}]{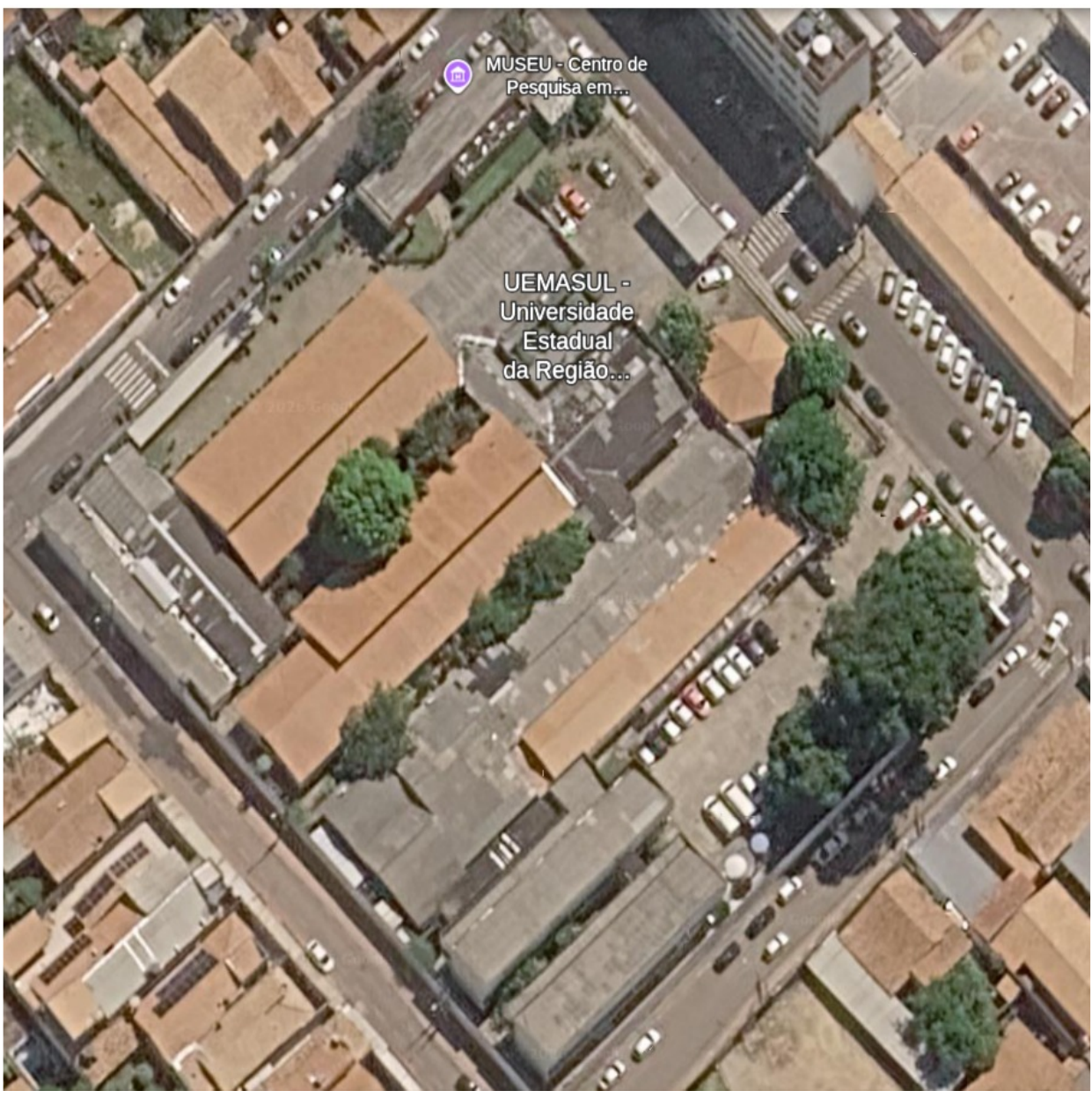}
		\caption{Aerial view of the UEMASUL headquarters. Source: Google Earth (2026) \cite{goo}. Map data.: Google, Maxar Technologies..}
		\label{1}
	\end{figure}
	
	Throughout the academic period, weekly meetings were held to develop this work, emphasizing discussions of issues related to climate change, relationships between society and nature, and the impacts of human activities on the environment. This approach considers environmental issues to be complex and to involve different dimensions, and therefore not to be treated in a fragmented manner within the educational process \cite{reis}.
	
	In this sense, the proposal sought to establish a relationship between physical knowledge and socio-environmental issues, using the greenhouse effect as a contextual element. Discussion of climate change offers possibilities for Physics teaching precisely because it enables the articulation of physical concepts, particularly those related to Thermodynamics, with environmental, social, and cultural issues \cite{magalhaes, reis}.
	
	\section{Bibliographic Review and Prototype Adaptation}
	
	The construction of the experimental apparatus was preceded by a bibliographic review of experimental proposals related to Physics education and environmental issues, particularly those using empirical models to discuss heating processes and the greenhouse effect. The studies consulted were used as references for defining the operating principles of the experiment, organizing the comparison systems, and selecting materials that could be used in constructing the prototype \cite{junges}.
	
	Based on the analysis of the experimental proposals consulted, the possibilities of reproducing and adapting the experiments were considered according to the materials available for this research. Aspects such as ease of assembly, material availability, low cost, portability, and the possibility of using the apparatus in teaching and science communication activities were considered.
	
	The prototype developed in this study does not correspond to an exact reproduction of a previously presented experiment. It is a methodological adaptation based on the references consulted, maintaining certain principles present in the experimental proposals analyzed while modifying the physical structure, materials, and organization of the system according to the conditions available for the investigation.
	
	One of the main adaptations consisted of using a glass aquarium as the confinement structure of the experimental system. This material was selected not only because of its availability, transparency, resistance, and ease of handling, but mainly because it allowed direct observation of the internal components of the experiment during its operation. In addition, the aquarium was used as a simplified representation of planet Earth, allowing a didactic relationship to be established between the experimental system and the terrestrial environment, in which solar radiation, the surface, and the atmosphere interact. Thus, the aquarium served as the structural element of the simulator, allowing the formation of a "confined system" and its comparison with an "open control system exposed to the environment."
	
\begin{figure}[htb!]
		\includegraphics[{angle=90,height=5.0cm,angle=270,width=5.0cm}]{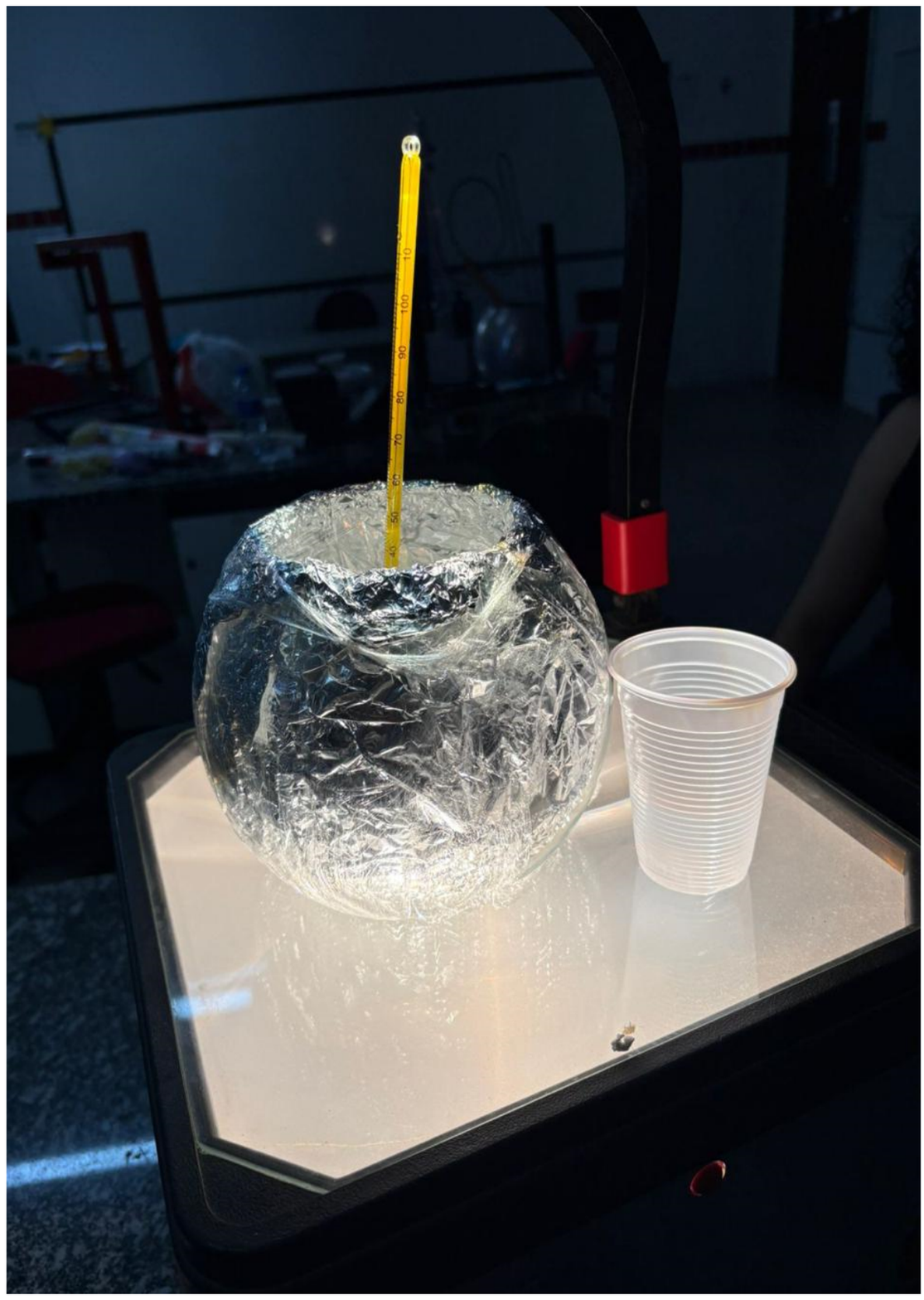}
		\caption{Our prototype}
		\label{2}
	\end{figure}

	The selection of the remaining materials was also based on the references consulted and the need to construct a low-cost and easily reproducible apparatus. Accessible materials were used, including disposable cups, water, thermometers, transparent PVC plastic film, black paint, and aluminum foil (see Fig.~\ref{2}).
	
	The adaptation of the materials and experimental configuration was therefore based on the experimental proposals consulted while taking into account the specific conditions of this investigation. This strategy made it possible to transform the experimental principles identified in the reference studies into an original prototype suitable for the material and pedagogical conditions of the project.

	\subsection{Science Awareness and Outreach Activities}

	The awareness activities aimed to extend the reach of the proposal beyond the formal educational environment, using the experimental apparatus as a science communication resource and as a means of bringing Physics knowledge closer to the community.
	
\begin{figure}[htb!]
		\includegraphics[{angle=90,height=6.5cm,angle=270,width=6.5cm}]{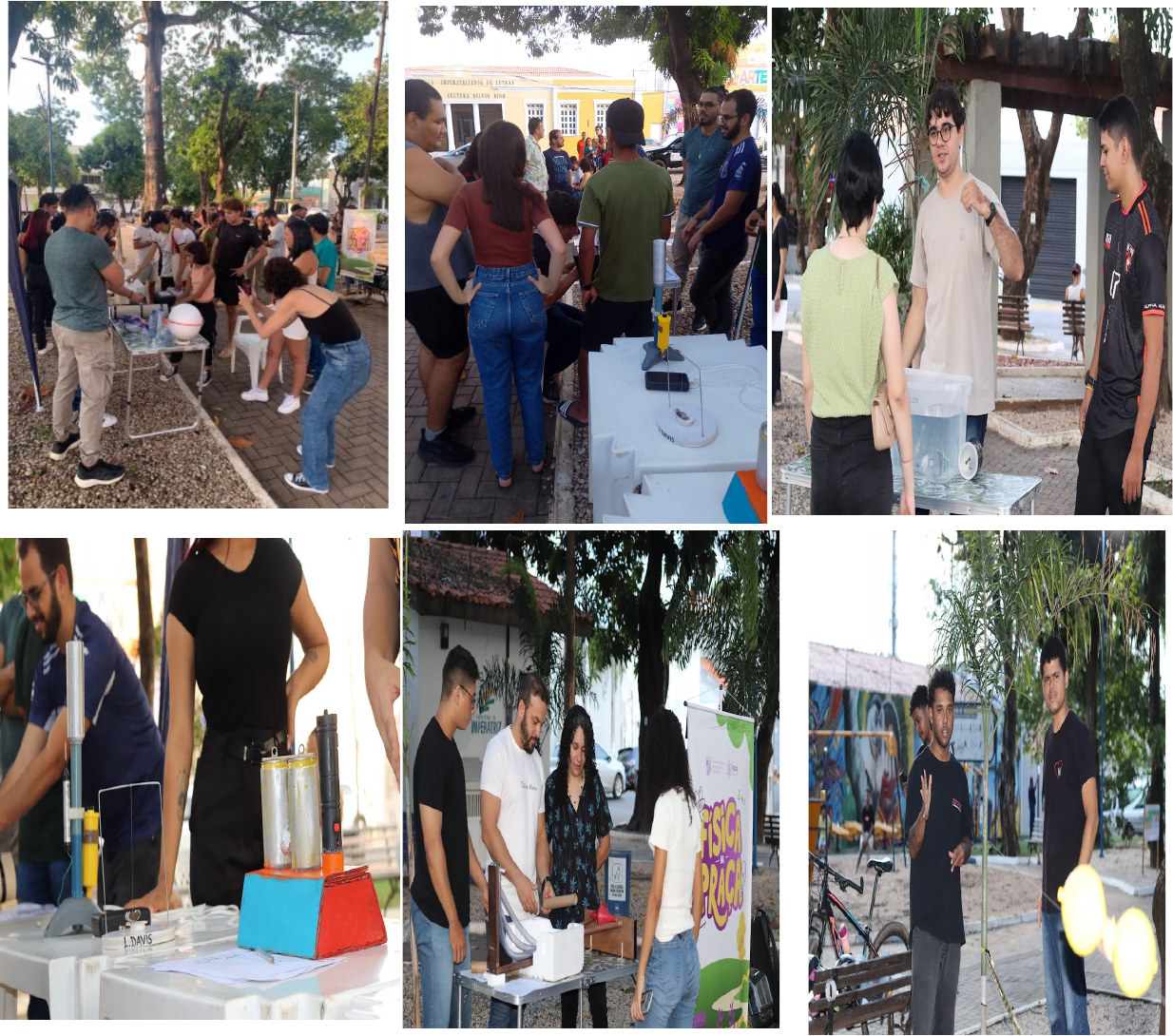}
		\includegraphics[{angle=90,height=4.5cm,angle=270,width=4.5cm}]{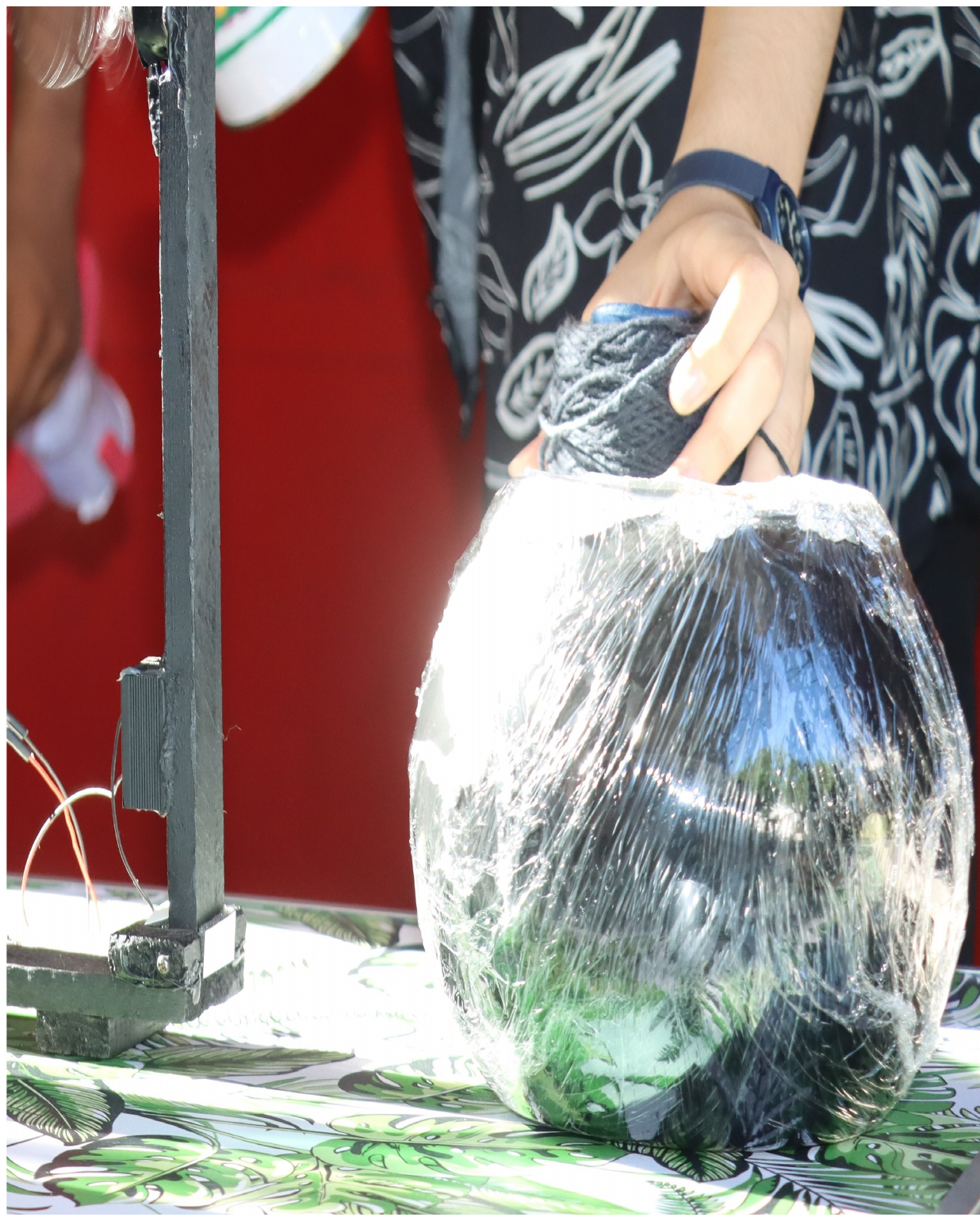}
		\caption{Science outreach activities}
		\label{3}
	\end{figure}

	The simulator was used in activities conducted in high-traffic public spaces in the city of Imperatriz, Maranhão (see Fig.~\ref{3}), particularly during Saturday interventions in the science outreach project "Physics in the Public Square" \cite{marinho}. These activities included didactic explanations and discussions with participants about the physical principles related to heating, thermal energy transfer, and the experimental representation used in the study.
	
	The use of experimental activities in this context sought to promote a connection between scientific knowledge and everyday situations, allowing physical concepts to be discussed based on a phenomenon directly observed by participants. This interaction between Physics and environmental issues has the potential to broaden understanding of phenomena and encourage a more reflective approach to socio-environmental issues \cite{reis, 3}.
	
	The proposal also considered that methodological approaches can significantly contribute to discussions of environmental issues, going beyond the boundaries of the school environment and reaching different social spaces through science communication and education activities. Thus, the methodology developed enabled the experiment to be used not only as a didactic resource for Physics teaching but also as a science communication tool, allowing scientific knowledge and environmental awareness to be disseminated in different social settings.
	
	\subsection{Materials and Experimental Apparatus}
	
	To construct the thermal simulator, low-cost and easily obtainable materials were used, with the aim of enabling the experimental practice to be reproduced in different educational contexts. The materials were selected based on the experimental proposals consulted and subsequently adapted to the conditions available for this research \cite{junges}.
	
	The physical structure of the prototype consisted of the following materials:
	
	\begin{itemize}
		\item 01 glass aquarium measuring $15~\text{cm} \times 18~\text{cm}$;
		\item black paint applied to the bottom and inner walls of the aquarium;
		\item sheets of aluminum foil used as an internal reflective coating;
		\item transparent plastic PVC film used to seal the upper part of the aquarium;
		\item 02 disposable cups with a capacity of $200~\text{mL}$, each containing $100~\text{mL}$ of water;
		\item 02 glass liquid thermometers with red expansion media.
	\end{itemize}
	
	The final apparatus configuration was established to enable comparison between two systems simultaneously subjected to the same solar radiation exposure conditions. The first system, referred to as the "confined system" or simulator, consisted of the aquarium containing a water container and a thermometer. The second, referred to as the open or control system, consisted of another water container and thermometer, remaining directly exposed to the local atmospheric conditions.
	
	\subsection{Experimental Procedure}
	
	The experimental apparatus was installed in an open area of Campus I of the Universidade Estadual da Região Tocantina do Maranhão (UEMASUL), in Imperatriz, Maranhão (see Fig.~\ref{1}), where it remained exposed to direct solar radiation.
	
	The "confined" experimental system consisted of the previously prepared aquarium containing a cup with $100~\text{mL}$ of water and an immersed thermometer. After the components were inserted, the upper opening of the aquarium was sealed with PVC plastic film, establishing the confinement condition used for the simulator.
	
\begin{figure}[htb!]
		\includegraphics[{angle=90,height=5.0cm,angle=270,width=5.0cm}]{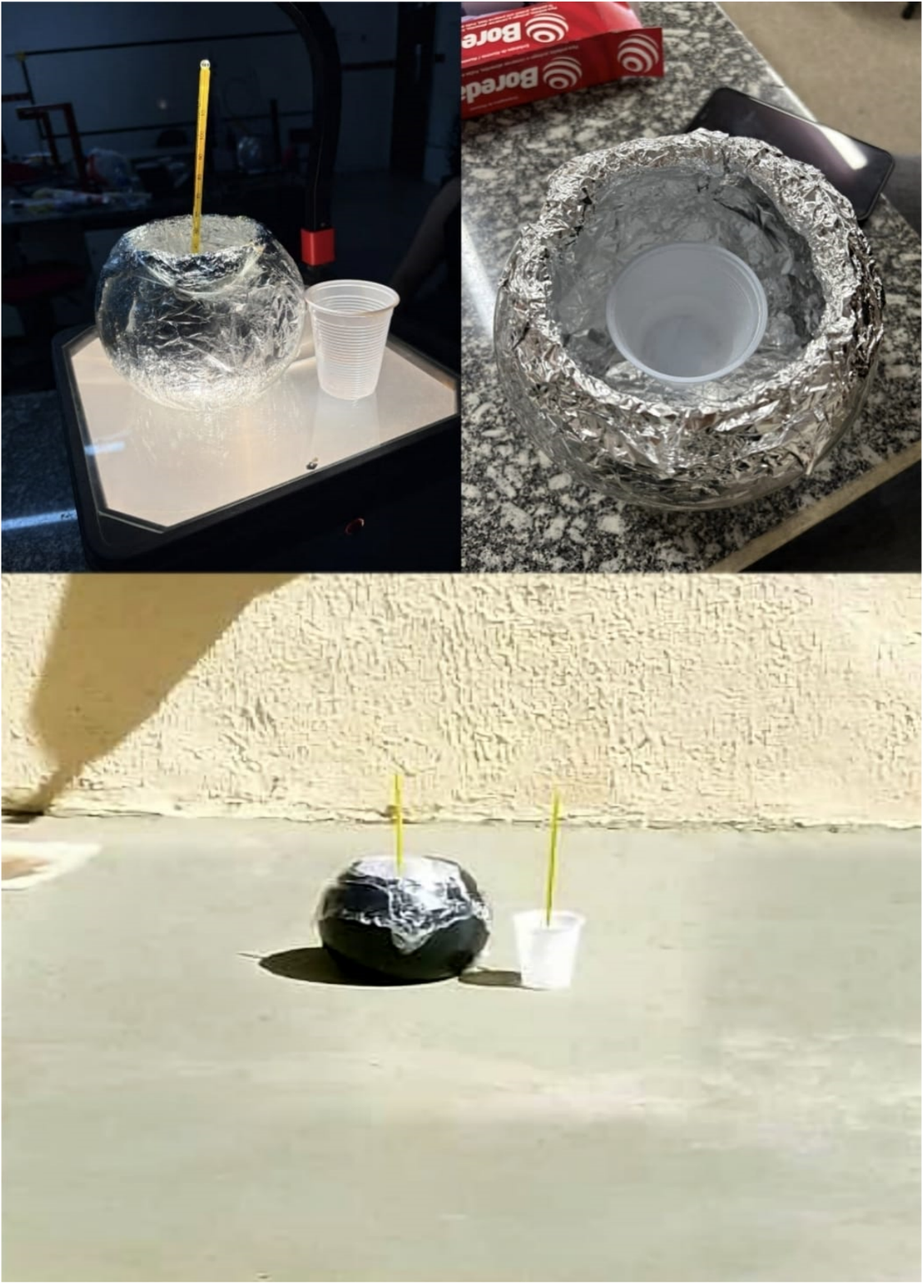}
		\caption{Date collection equipment and location}
		\label{5}
	\end{figure}

\begin{figure}[htb!]
		\includegraphics[{angle=90,height=5.0cm,angle=270,width=5.0cm}]{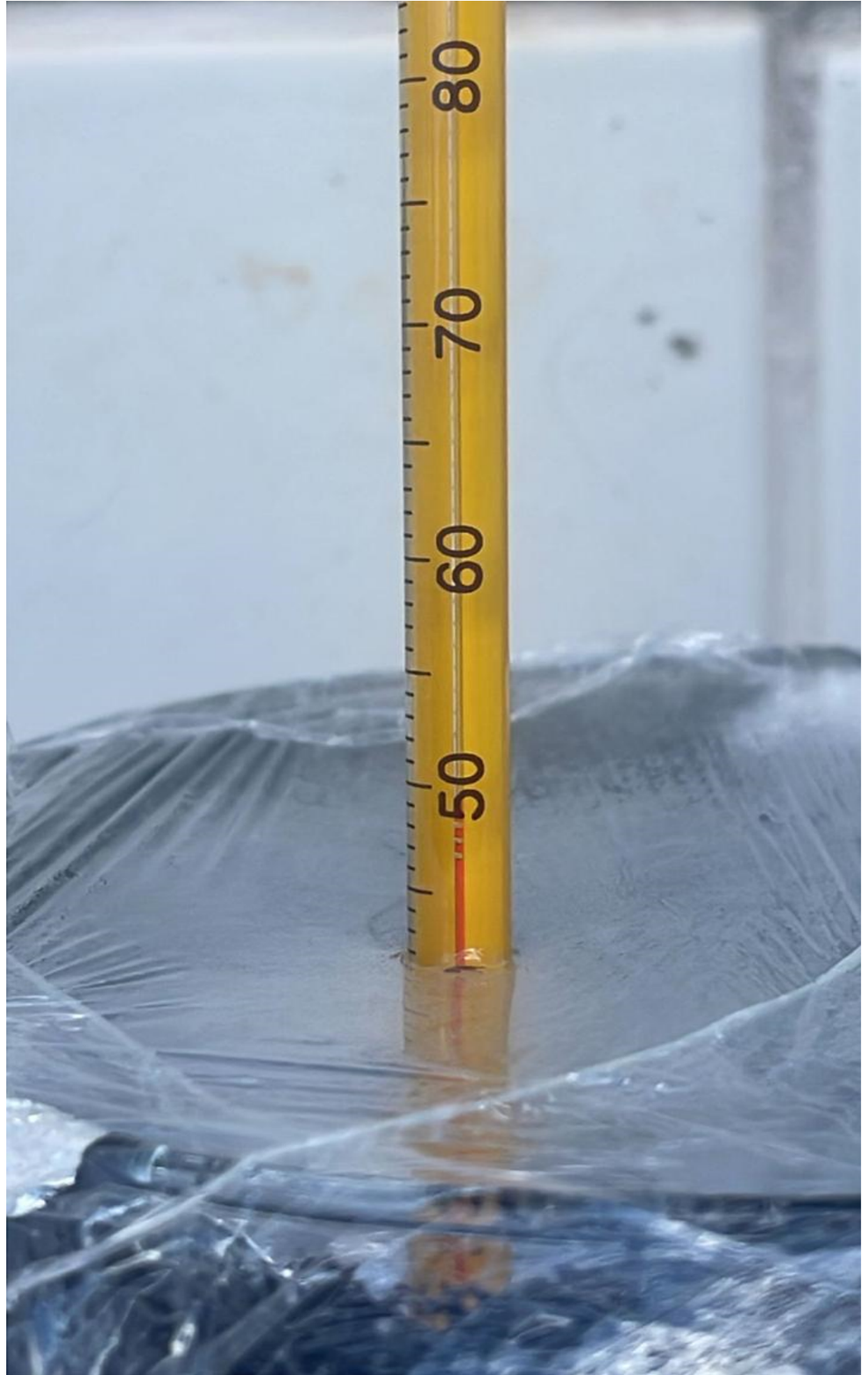}
		\includegraphics[{angle=90,height=5.0cm,angle=270,width=5.0cm}]{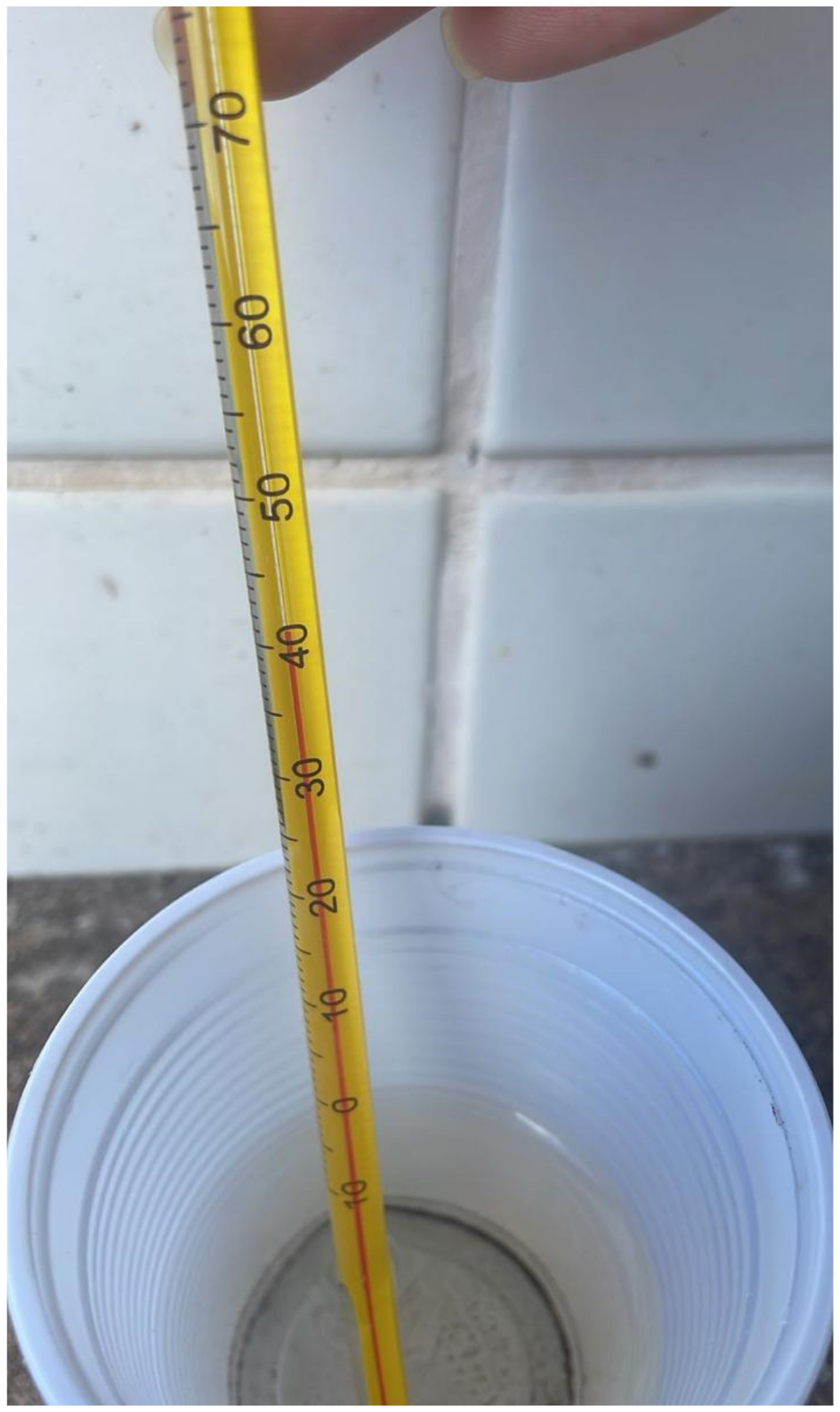}
		\caption{Some measurements}
		\label{6}
	\end{figure}

	Simultaneously, an open control system was established, consisting of a second cup containing $100~\text{mL}$ of water and the second thermometer. This system was positioned close to the simulator and remained directly exposed to local atmospheric conditions. In Figs.~\ref{5} and \ref{6}, the locations where the apparatuses were exposed during the data-collection period and some of the measurements obtained can be observed.
	
	The use of two systems simultaneously subjected to the exposure conditions was intended to allow comparison between the temperatures recorded inside the simulator and those observed in the open system. Thus, the objective was to evaluate the difference in thermal behavior between the two experimental conditions.

The adopted configuration was based on the references consulted but incorporated original modifications to the physical structure, materials employed, and organization of the experimental system. These modifications were made to adapt the prototype to the conditions available for the research and science outreach activities.
	
	\subsection{Date Collection and Organization}
	
	Data were systematically collected over 11 consecutive days, from June 16 to June 26, 2026. The observation period was established within this interval because of the winter solstice, as previously mentioned, which served as our temporal reference.
	
	On each observation day, five temperature measurements were performed at approximately 30-minute intervals. Measurements were taken at the previously established times of 13h24, 13h55, 14h25, 14h55, and 15h29.
	
	At each time, the temperatures corresponding to the aquarium, representing the simulator, and the open system, used as the control, were recorded simultaneously. Thus, temperature time series were obtained for both experimental conditions.
	
	The collected data were subsequently organized into tables, enabling comparison between the temperatures of the "confined" and open systems throughout the observation period. The difference between the recorded values was used as a parameter for analyzing the thermal behavior of the systems.
	
	Considering that the prototype constitutes a simplified experimental representation, the results were interpreted within the limits of the configuration used. The experiment was not intended to reproduce the full complexity of the Earth's climate system but rather to enable the observation and discussion of thermal processes under a controlled experimental situation. This limitation is important given the complexity associated with climate phenomena and the radiative-transfer processes involved in the greenhouse effect \cite{4}.

	\section{Results and Discussion}
	
	The experiment was conducted over 11 consecutive days, from June 16 to June 26, 2026, with the objective of monitoring temperature variation inside a low-cost greenhouse-effect simulator and comparing it with the temperature of the external environment. For the analysis, the internal ($T_{\text{inside}}$) and external ($T_{\text{outside}}$) temperatures were considered simultaneously, and the temperature difference between the two environments was defined as
	
	\begin{equation}
		\Delta T(t) = T_{\text{inside}}(t) - T_{\text{outside}}(t)
	\end{equation}
	
	The use of $\Delta T$ makes it possible to evaluate the thermal difference between the interior and exterior of the simulator, considering the environmental conditions present during each observation period.
	
	The results presented in Table~\ref{tab:resumo_diario} show that, on all analyzed days, the mean temperature inside the simulator was higher than the mean external temperature. Considering the entire experimental period, the mean external temperature was $38.58^{\circ}\text{C}$, whereas the mean internal temperature reached $45.91^{\circ}\text{C}$. Thus, the mean temperature difference was
	
	\begin{equation}
		\overline{\Delta T} = 7.33^{\circ}\text{C}.
	\end{equation}
	
	\begin{table}[htbp]
		\centering
		\caption{Daily statistical summary of temperatures recorded inside and outside the simulator from June 16 to June 26, 2026.}
		\label{tab:resumo_diario}
		\resizebox{0.95\textwidth}{!}{%
			\begin{tabular}{ccccccc}
				\toprule
				\textbf{Date} &
				\textbf{$\overline{T}_{\text{outside}}$ ($^\circ$C)} &
				\textbf{$\overline{T}_{\text{inside}}$ ($^\circ$C)} &
				\textbf{$\overline{\Delta T}$ ($^\circ$C)} &
				\textbf{$T_{\text{max out}}$ ($^\circ$C)} &
				\textbf{$T_{\text{max in}}$ ($^\circ$C)} &
				\textbf{$\Delta T_{\text{max}}$ ($^\circ$C)} \\
				\midrule
				16/06 & 36,00 & 41,80 & 5,80 & 39,0 & 50,0 & 11,0 \\
				17/06 & 38,25 & 44,75 & 6,50 & 39,0 & 48,0 & 9,0 \\
				18/06 & 37,75 & 45,00 & 7,25 & 40,0 & 48,0 & 10,0 \\
				19/06 & 38,50 & 43,50 & 5,00 & 41,0 & 49,0 & 8,0 \\
				20/06 & 36,75 & 43,00 & 6,25 & 38,0 & 48,0 & 10,0 \\
				21/06\textsuperscript{*} & 38,00 & 45,25 & 7,25 & 40,0 & 47,0 & 9,0 \\
				22/06 & 40,75 & 49,50 & 8,75 & 42,0 & 55,0 & 13,0 \\
				23/06 & 41,25 & 50,50 & 9,25 & 42,0 & 54,0 & 12,0 \\
				24/06 & 40,50 & 48,75 & 8,25 & 42,0 & 52,0 & 10,0 \\
				25/06 & 40,00 & 49,50 & 9,50 & 41,0 & 54,0 & 13,0 \\
				26/06 & 37,25 & 44,50 & 7,25 & 40,0 & 48,0 & 11,0 \\
				\midrule
				\textbf{Overall Average} &
				\textbf{38,58} &
				\textbf{45,91} &
				\textbf{7,33} &
				\textbf{40,36} &
				\textbf{50,27} &
				\textbf{10,36} \\
				\bottomrule
			\end{tabular}%
		}
		
		\vspace{2mm}
		
		\raggedright
		\footnotesize
		\textsuperscript{*} Winter Solstice in the Southern Hemisphere.
		
	\end{table}
	
This result indicates that the conditions established inside the simulator produced a thermal condition distinct from that observed in the external environment. However, the temperature difference did not remain constant throughout the experimental period, varying among the different observation days.

The lowest mean value of $\Delta T$ was recorded on June 19, at $5.00^{\circ}\text{C}$, whereas the highest mean value occurred on June 25, reaching $9.50^{\circ}\text{C}$. Regarding the maximum instantaneous values, the largest gradients were observed on June 22 and June 25, both with $\Delta T_{\text{max}}=13.0^{\circ}\text{C}$. On June 22, the maximum temperature recorded inside the simulator was $55.0^{\circ}\text{C}$, while the maximum external temperature was $42.0^{\circ}\text{C}$.

	\begin{figure}[htbp]
		\centering
		\begin{tikzpicture}
			\begin{axis}[
				width=0.9\linewidth,
				height=7cm,
				ybar,
				bar width=14pt,
				xlabel={Date},
				ylabel={Maximum Difference $\Delta T_{\text{max}}\,(^{\circ}\text{C})$},
				symbolic x coords={16/06,17/06,18/06,19/06,20/06,21/06,22/06,23/06,24/06,25/06,26/06},
				xtick=data,
				nodes near coords,
				nodes near coords align={vertical},
				ymin=0,
				ymax=16,
				ymajorgrids=true,
				grid style=dashed,
				x tick label style={rotate=45, anchor=east},
				fill=deltacolor!80!white,
				draw=deltacolor!40!black
				]
				
				\addplot coordinates {
					(16/06,11)
					(17/06,9)
					(18/06,10)
					(19/06,8)
					(20/06,10)
					(21/06,9)
					(22/06,13)
					(23/06,12)
					(24/06,10)
					(25/06,13)
					(26/06,11)
				};
				
			\end{axis}
		\end{tikzpicture}
		
		\caption{Maximum temperature difference ($\Delta T_{\text{max}}$) recorded by the simulator on each sampling day.}
		\label{fig:delta_max_barras}
	\end{figure}
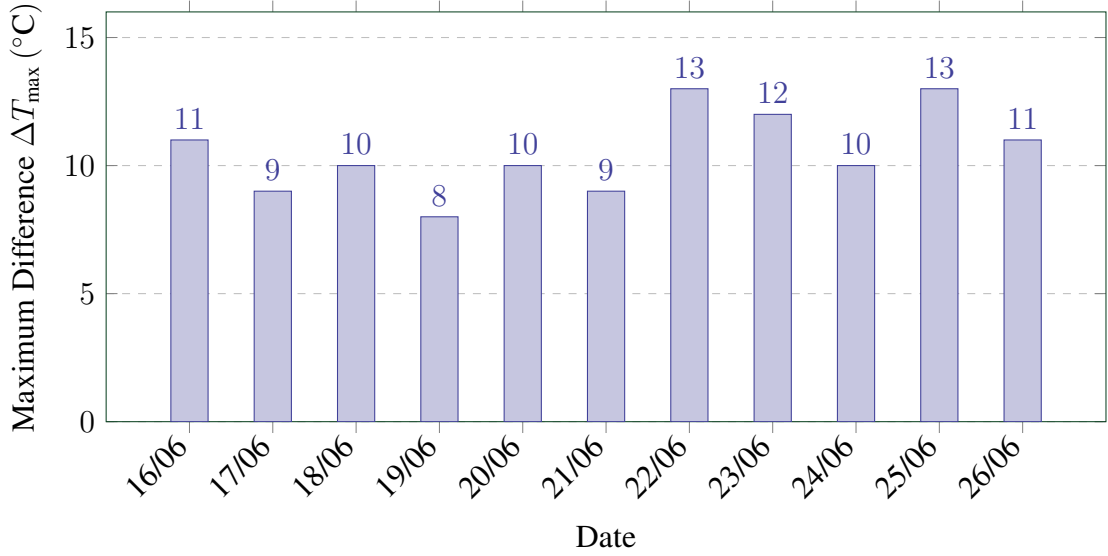

	Figure~\ref{fig:delta_max_barras} presents the distribution of the maximum $\Delta T$ values over the 11 observation days and shows that the highest values of $\Delta T_{\text{max}}$ occurred on June 22 and June 25, both at $13.0^{\circ}\text{C}$, followed by $12.0^{\circ}\text{C}$ on June 23. In contrast, the lowest maximum value was recorded on June 19, at $8.0^{\circ}\text{C}$. This variation demonstrates that the thermal gradient produced by the simulator is not a fixed quantity but depends on the conditions to which the system was subjected on each day.
	
	The difference observed between the internal and external temperatures can be interpreted based on the energy balance of the system. Incident solar radiation supplies energy to the simulator, part of which is absorbed by the surfaces present in the system and subsequently converted into thermal energy. Heated materials then emit thermal radiation, predominantly in the infrared region, while heat-transfer processes by conduction and convection also occur.
	
	This behavior is related to the general principle of energy conservation applied to thermal systems. In the Earth system, energy originating from solar radiation is absorbed by the surface and atmosphere and subsequently redistributed and emitted through different mechanisms. The balance between incoming energy and energy loss determines the thermal evolution of the system \cite{NASA2009, ipcc}.
	
	It is important, however, to distinguish between the operation of the simulator used in this experiment and the physical mechanism of the atmospheric greenhouse effect. In the Earth's atmosphere, the greenhouse effect is primarily associated with the absorption and emission of infrared radiation by radiatively active gases such as water vapor, carbon dioxide, and methane. These gases absorb specific portions of the infrared radiation emitted by the surface and subsequently emit radiation in different directions, including a component directed toward the surface \cite{NASA2009, NASA2023}.
	
	In the experimental simulator, on the other hand, additional mechanisms associated with air confinement and the physical characteristics of the container can modify heat exchange with the external environment. Therefore, the increase in temperature observed inside the simulator should not be attributed exclusively to the retention of infrared radiation. The value of $\overline{\Delta T}=7.33^{\circ}\text{C}$ represents the thermal difference produced under the established experimental conditions and does not constitute a direct measurement of the intensity of the atmospheric greenhouse effect.
	
	The temporal evolution of temperatures provides additional information about the system's behavior. As shown in Table~\ref{tab:resumo_horario}, at 13:55 the mean external temperature was $37.73^{\circ}\text{C}$, while the mean internal temperature was $42.27^{\circ}\text{C}$, resulting in a mean difference of $4.55^{\circ}\text{C}$. At 14h25, the values were $39.18^{\circ}\text{C}$ and $46.82^{\circ}\text{C}$, respectively, causing the thermal gradient to increase to $7.64^{\circ}\text{C}$. At 14h55, the difference reached $8.55^{\circ}\text{C}$, while at 15h29 the highest mean value, $9.27^{\circ}\text{C}$, was recorded.
	
	\begin{table}[htbp]
		\centering
		\caption{Time-averaged values consolidated by sampling time over the entire experimental period.}
		\label{tab:resumo_horario}
		
		\begin{tabular}{cccc}
			\toprule
			\textbf{Time} &
			$\mathbf{\overline{T}_{\text{outside}}\,(^{\circ}\text{C})}$ &
			$\mathbf{\overline{T}_{\text{inside}}\,(^{\circ}\text{C})}$ &
			$\mathbf{\overline{\Delta T}\,(^{\circ}\text{C})}$ \\
			\midrule
			13:24\textsuperscript{\dag} & 28,00 & 28,00 & 0,00 \\
			13:55 & 37,73 & 42,27 & 4,55 \\
			14:25 & 39,18 & 46,82 & 7,64 \\
			14:55 & 39,27 & 47,82 & 8,55 \\
			15:29 & 39,09 & 48,36 & 9,27 \\
			\bottomrule
			\multicolumn{4}{l}{\footnotesize
				\textsuperscript{\dag}Measurement performed only on June 16
				(initial calibration point).}
		\end{tabular}
	\end{table}
	
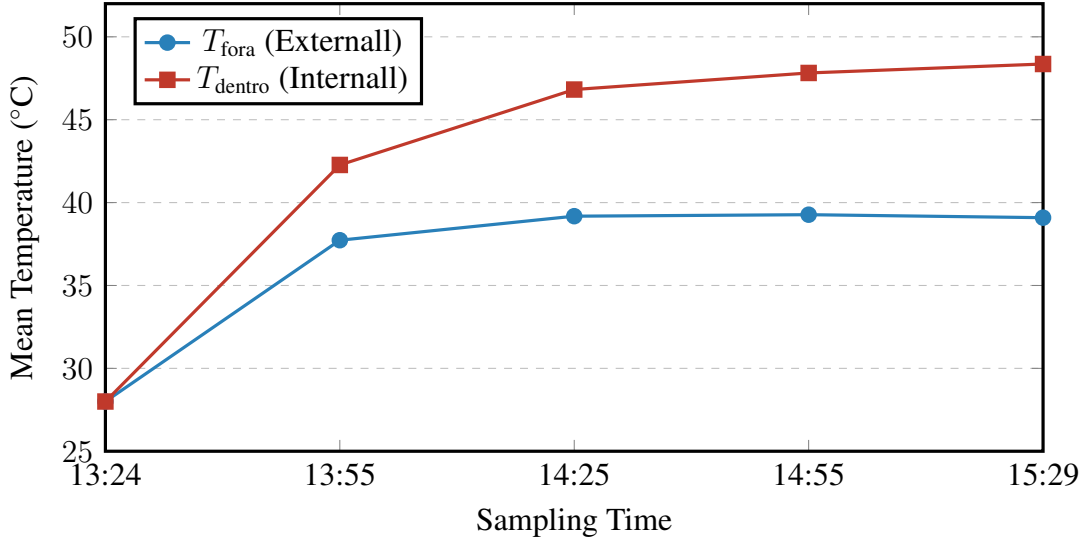
\begin{figure}[htbp]
		\centering
		
		\begin{tikzpicture}
			\begin{axis}[
				width=0.85\linewidth,
				height=7.5cm,
				xlabel={Sampling Time},
				ylabel={Mean Temperature ($^{\circ}\text{C}$)},
				xmin=1,
				xmax=5,
				ymin=25,
				ymax=52,
				xtick={1,2,3,4,5},
				xticklabels={13:24,13:55,14:25,14:55,15:29},
				legend pos=north west,
				ymajorgrids=true,
				grid style=dashed,
				line width=1.2pt,
				mark size=2.5pt
				]
				
				\addplot[color=tempfora, mark=*] coordinates {
					(1,28.00)
					(2,37.73)
					(3,39.18)
					(4,39.27)
					(5,39.09)
				};
				
				\addlegendentry{$T_{\text{fora}}$ (Externall)}
				
				\addplot[color=tempdentro, mark=square*] coordinates {
					(1,28.00)
					(2,42.27)
					(3,46.82)
					(4,47.82)
					(5,48.36)
				};
				
				\addlegendentry{$T_{\text{dentro}}$ (Internall)}
				
			\end{axis}
		\end{tikzpicture}
		
		\caption{Mean temporal evolution of external ($T_{\text{outside}}$) and internal ($T_{\text{inside}}$) temperatures throughout the afternoon.}
		\label{fig:curva_temporal}
	\end{figure}

	Figure~\ref{fig:curva_temporal} illustrates this behavior, showing that the internal temperature remained higher than the external temperature during the considered sampling times. It can also be observed that the difference between the two curves progressively increased throughout the afternoon.
	
	One possible interpretation of this behavior is related to the accumulation of thermal energy within the system. During exposure to solar radiation, part of the incoming energy is absorbed by the materials present in the simulator and converted into internal energy. As long as there is a net gain of energy, the system temperature tends to increase. Simultaneously, energy is transferred to the external environment through different mechanisms, such that the temperature observed at any given instant results from the balance between incoming energy and energy losses.
	
	This interpretation is consistent with the concept of the Earth's surface energy balance. According to the IPCC, the surface energy balance involves radiative and non-radiative components, including solar and thermal radiation, sensible and latent heat fluxes, and heat flux into the soil \cite{ipcc}. Similarly, in the experimental system, the internal temperature results from the interaction among processes involving energy input, energy storage, and energy transfer.
	
	The variation in $\Delta T$ among different days also demonstrates that the simulator's thermal behavior depends on the exposure conditions during each measurement period. Although the experimental period included the winter solstice in the Southern Hemisphere on June 21, 2026, the data obtained do not allow a causal relationship between the solstice and the observed temperature values to be established in isolation.
	
	Likewise, direct measurements of solar irradiance, cloud cover, wind speed, or relative humidity were not performed. Therefore, based solely on the temperature data, it is not possible to state that the largest gradients observed on June 22, 23, and 25, 2026, were caused exclusively by greater solar radiation exposure.
	
	This consideration is important because solar energy constitutes the primary energy source of the climate system, but the amount actually absorbed by the surface depends on different atmospheric and surface conditions. Earth's energy balance simultaneously involves incoming radiation, reflected radiation, emitted thermal radiation, and non-radiative energy-transfer processes \cite{NASA2009, ipcc}. Therefore, interpretation of the experimental results must consider the existence of multiple factors capable of influencing the recorded temperature.
	
	It should also be emphasized that the temperature difference observed in the experiment cannot be directly compared with the increase in global mean temperature associated with changes in the atmospheric greenhouse effect. The IPCC characterizes the effects of changes in atmospheric composition through changes in the energy balance of the Earth system and the concept of radiative forcing, which involves energy fluxes at the planetary scale \cite{ipcc}. The experiment developed in this study has a much more restricted spatial, temporal, and physical scale and is therefore essentially demonstrative and educational in purpose.
	
	In this sense, the main experimental result is the systematic observation that the temperature inside the simulator was higher than the external temperature during the analyzed period. The mean difference of $7.33^{\circ}\text{C}$ and maximum values of up to $13.0^{\circ}\text{C}$ demonstrate that the experimental configuration was capable of producing a measurable thermal difference between the two environments.
	
	The temporal evolution also presents a relevant result. The mean temperature difference increased from $4.55^{\circ}\text{C}$ at 13:55 to $9.27^{\circ}\text{C}$ at 15h29. This behavior demonstrates that analysis of a thermal system should consider not only instantaneous temperature values but also the temporal evolution of the energy balance.
	
	Therefore, the results demonstrate that the low-cost simulator was capable of producing and recording a significant thermal difference relative to the external environment. The observed behavior can be used as a simplified experimental representation of heating and energy-transfer processes associated with exposure to solar radiation. However, the analogy with the atmospheric greenhouse effect should be made cautiously, since the heating observed in the container results from a combination of radiative and non-radiative mechanisms, whereas the atmospheric greenhouse effect is directly related to the radiative properties of greenhouse gases.
	
	Thus, within the limitations of the experimental apparatus, the results provide a quantitative and accessible approach for discussing concepts related to energy balance, heat transfer, and the thermal response of a system subjected to solar radiation. The use of a low-cost simulator also makes it possible to bring these concepts closer to a concrete experimental situation, allowing the formation and evolution of a thermal gradient over time to be directly observed.
	
	\section{Conclusion}
	
	The present study made it possible to experimentally investigate the thermal behavior of a low-cost simulator subjected to solar radiation exposure, establishing a comparison with a control system exposed to local atmospheric conditions. Conducting the experiment over 11 days allowed the evolution of temperatures to be monitored and the differences in behavior between the two systems to be quantitatively observed.
	
	The results demonstrated that, during the analyzed period, the mean temperature inside the simulator was higher than the mean external temperature, with mean values of 45.91 °C and 38.58 °C, respectively, resulting in a mean temperature difference of 7.33 °C. Maximum temperature differences of up to 13.0 °C were also observed. These results indicate that the experimental configuration used was capable of producing and recording a measurable thermal difference between the two analyzed conditions.
	
	The temporal analysis further showed that the temperature difference increased throughout the afternoon, from 4.55 °C at 13h55 to 9.27 °C at 15h29. This behavior reinforces the importance of considering the temporal evolution of the system, since the observed temperature results from the balance among incoming energy, energy storage, and different forms of heat transfer.
	
	Although the observed behavior presents a didactic analogy with processes related to the heating of the Earth's surface, the results should not be interpreted as a direct measurement of the atmospheric greenhouse effect or global warming. The simulator has a simplified physical configuration in which air confinement and the characteristics of the materials used influence energy exchanges. Unlike the atmospheric greenhouse mechanism, which is related to the radiative properties of greenhouse gases, the heating observed in the apparatus results from a combination of different energy-transfer mechanisms.
	
	The variation observed among the different days also demonstrates that the system's thermal behavior is related to the exposure conditions during each observation period. Since direct measurements of solar irradiance, cloud cover, wind speed, or relative humidity were not performed, it is not possible to attribute these variations to a single environmental factor. This limitation reinforces the need to interpret the results while considering the complexity of the processes involved in the thermal dynamics of the environment. As a perspective for future research, we propose expanding this methodology to different geographic regions, enabling comparative analysis of micrometeorological variables—such as solar irradiance, relative humidity, wind speed, and ambient temperature—through the integration of digital sensors.
	
	In this sense, the experiment proved suitable as a resource for addressing fundamental concepts of Physics and Environmental Sciences, particularly those related to energy balance, temperature, thermal gradients, and heat-transfer processes. The use of accessible materials also expands the possibilities of reproducing the activity in different educational contexts, bringing scientific concepts closer to situations observable in everyday life.
	
	In addition to its experimental dimension, the activity demonstrated potential for the initial training of Physics teachers by articulating physical knowledge, environmental issues, experimental investigation, and science communication. The use of the apparatus in extension activities allowed this approach to be expanded beyond the formal educational environment, promoting the communication of scientific concepts and strengthening the relationship between the university and society.
	
	Finally, the results indicate that low-cost experiments can constitute relevant tools for science education when accompanied by theoretical foundations, systematic procedures for data collection and analysis, and, above all, an explicit discussion of the limitations of the analogies and models employed. Thus, rather than reproducing the complex terrestrial climate system, the simulator developed in this study made it possible to transform an everyday phenomenon into an object of investigation, contributing to the construction of knowledge about Physics, the environment, and climate science in the education of future teachers.
	

	\section*{Author Contributions}
	
	All authors contributed equally to all stages of the manuscript. 
	
	All authors have read and agreed to the published version of the manuscript.

	\section*{Acknowledgments}
	
	The authors would like to thank the CAPES, FAPEMA and UEMASUL  for their financial support. 
	
	To the public who authorized the use of the images for scientific dissemination purposes.

\end{document}